\documentclass[aps,prd,twocolumn,showpacs,groupedaddress]{revtex4}
\usepackage{graphicx}
\usepackage{latexsym}
\usepackage{subfigure}

\begin{document}
\title{Experimental Requirements to Determine the Neutrino Mass Hierarchy Using Reactor Neutrinos}

\author{Liang Zhan, Yifang Wang, Jun Cao, Liangjian Wen}

\affiliation{Institute of High Energy Physics, Chinese Academy of
Sciences, Beijing 100049, China}

\begin{abstract}
This paper presents experimental requirements to determine the
neutrino mass hierarchy using reactor neutrinos. The detector
shall be located at a baseline around 58~km from the reactor(s) to
measure the energy spectrum of electron antineutrinos
($\overline{\nu}_e$) precisely. By applying Fourier cosine and
sine transform to the L/E spectrum, features of the neutrino mass
hierarchy can be extracted from the $|\Delta{m}^2_{31}|$ and
$|\Delta{m}^2_{32}|$ oscillations. To determine the neutrino mass
hierarchy above 90\% probability, requirements to the baseline,
the energy resolution, the energy scale uncertainty, the detector
mass and the event statistics are studied at different values of
$\sin^2(2\theta_{13})$.
\end{abstract}

\pacs{13.15.+g, 14.60.Pq, 14.60.Lm}
\maketitle

Neutrino physics has undergone a revolution over the last decade
and reaches now to an era of precision measurement for the
neutrino oscillation parameters. However, $\theta_{13}$,
CP-violating phase and the sign of $\Delta m^2_{32}$ (mass
hierarchy) are still undetermined. Usually normal hierarchy (NH)
is defined as $\Delta m^2_{32}>0$ and inverted hierarchy (IH) is
defined as $\Delta m^2_{32}<0$. Accelerator neutrino experiments
such as Nova \cite{Mena:2005ri, Ayres:2004js, Mena:2006uw} and
T2KK \cite{Hagiwara:2006vn} have the potential to determine the
mass hierarchy using the matter effect of neutrinos at long
baselines. There are also discussions to precisely measure the
distortions of the reactor neutrino energy spectrum at an
intermediate baseline ($40-65~$km) \cite{Petcov:2001sy,
Choubey:2003qx}. A Fourier transform method was recently proposed
to enhance and visualize the features of mass hierarchy in the
frequency ($\Delta m^2$) spectrum \cite{Learned:2006wy}.

A new study \cite{Zhan:2008id} based on Fourier transformation
utilizing both the amplitude and phase information is presented
recently to enhance the features distinguishing the mass hierarchy
at a very small $\sin^2(2\theta_{13})$ value. This paper is
complimentary to Ref. \cite{Zhan:2008id} by taking into account
experimental details, such as the baseline, detector response
including energy resolution, energy scale uncertainty, and the
event statistics, etc. The study is based on Monte Carlo
simulation.

Taking into account the detector response, the reactor neutrino
$\overline{\nu}_e$ L/E spectrum F(L/E) becomes
\begin{eqnarray}
\label{eq:LoEspectrum}
 F(L/E') &=& \int R(E, E') F(L/E) \mathrm{d}E, \nonumber \\
 F(L/E) &=& \phi(E)\sigma(E) P_{ee}(L/E),
\end{eqnarray}
where $L$ is the baseline, $E$ is the actual $\overline{\nu}_e$
energy, $E'$ is the observed $\overline{\nu}_e$ energy taking into
account the detector response, and $R(E, E')$ represents the
detector response including effects such as the energy resolution
and energy scale. The reactor neutrino flux, $\phi(E)$, the
neutrino inverse beta reaction cross section with detector,
$\sigma(E)$, and the neutrino oscillation probability,
$P_{ee}(E)$, have all been described in Ref. \cite{Zhan:2008id}.
Here for completeness, we rewrite the $\overline{\nu}_e$ survival
probability $P_{ee}(E)$\cite{Bilenky:2001jq} as:

\begin{eqnarray}
\label{eq:Pee}
 P_{ee}(L/E) &=& 1 - P_{21} - P_{31} - P_{32}, \nonumber \\
 P_{21} &=& \cos^4(\theta_{13})\sin^2(2\theta_{12})\sin^2(\Delta_{21}),\nonumber\\
 P_{31} &=& \cos^2(\theta_{12})\sin^2(2\theta_{13})\sin^2(\Delta_{31}),\nonumber\\
 P_{32} &=&
 \sin^2(\theta_{12})\sin^2(2\theta_{13})\sin^2(\Delta_{32}),
\end{eqnarray}

The analytical formulas for Fourier cosine and sine transform
\cite{Zhan:2008id} are:
\begin{eqnarray}
\label{eq:FT}
FCT(\omega) = \int^{t_{max}}_{t_{min}}F(t) \cos (\omega t)\mathrm{d}t, \nonumber \\
FST(\omega) = \int^{t_{max}}_{t_{min}}F(t) \sin(\omega
t)\mathrm{d}t,
\end{eqnarray}
where $\omega$ is the frequency defined as $2.54 \Delta m^{2}$;
$t=L/E$ is the variable in L/E space, varying from $t_{min} =
L/E_{max}$ to $t_{max} = L/E_{min}$. In real experiments with a
set of discrete events, the integral can be changed to the
summation over all events as in the following:
\begin{eqnarray}
\label{eq:FTexp}
FST(\omega) = \sum^{N}_{i=1} \sin (\omega L/E'_{i}), \nonumber \\
FCT(\omega) = \sum^{N}_{i=1} \cos (\omega L/E'_{i}),
\end{eqnarray}
where $E'_{i}$ is the measured energy of individual events, and N
is the total number of events collected.

The actual experimental measurements of the neutrino energy
usually have two aspects of detector responses: energy resolution
and energy scale. The response of the detector due to energy
resolution can usually be described by a Gaussian function
$\frac{1}{\sqrt{2\pi}\sigma}exp(-\frac{(E'-E)^2}{2{\sigma_{E}}^2})$,
where $\sigma_{E}$ is the energy resolution. Since the neutrino
energy are usually measured by scintillators, the energy is
typically proportional to the number of photoelectrons, and the
error is dominated by the photoelectron statistics. Therefore the
neutrino energy resolution is proportional to $1/\sqrt{E_{vis}}$,
where $E_{vis} = E_{\nu}-0.8$ MeV is the neutrino visible energy
in the detector. Previous experiments typically have an energy
resolution of about $10\%/\sqrt{E_{vis}}$.   Different detectors
may have different forms of the energy scale uncertainty. For
simplicity, we take two possible cases, shift and
shrinking/expanding. It is modelled as formula $E'=(1+a)E+b$,
where $a$ and $b$ are parameters.

In this study, each Monte Carlo experiment generates a set of
$\overline{\nu}_e$ events by sampling F(L/E¡¯) spectrum with input
parameters $\{\sin^2\theta_{13}, L, \sigma_{E}, a, b\}$. The total
number of generated events determines the statistical error.
Default oscillation parameters are taken from Ref.
\cite{Zhan:2008id} and reproduced here in Table \ref{tab:default},
together with default input parameters to be studied in this
paper.

\begin{table}
\begin{tabular}{|c|c|c|c|}
  \hline
  $\Delta m^2_{21}$ & $|\Delta m^2_{32}|$ & $\sin^2\theta_{12}$ & $\sin^2\theta_{23}$ \\
  \hline
  $7.6\times 10^{-5}{\rm eV}^2$ & $2.4\times 10^{-3} {\rm eV}^2$ & 0.32 & 0.50 \\
  \hline
\end{tabular}
\begin{tabular}{|c|c|c|c|c|}
  \hline
  $\sin^2(2\theta_{13})$ & L & $\sigma_{E}$ & a & b \\
  \hline
  0.02 & 58~km & $3\%/\sqrt{E_{vis}}$ & 1\% & 0.01~MeV \\
  \hline
\end{tabular}
\caption{Default values for neutrino oscillation parameters and
other input parameters studied in this paper.} \label{tab:default}
\end{table}

In the following study, effects of input parameters to the mass
hierarchy are studied one by one, while the rest remain at the
default values. It is known that at the oscillation maximum of
$\Delta m^2_{12}$, corresponding to a baseline of about 58~km, the
sensitivity to the mass hierarchy is maximized. Hence the default
value of the baseline is set at 58~km. The mass hierarchy effects
in the reactor neutrino energy spectrum are mainly characterized
by ~$\Delta m^2_{21}/|\Delta m^2_{32}|$, which is only about 3\%.
Hence the required energy resolution shall be at the level of this
number and the default value in this study is set to be
$3\%/\sqrt{E_{vis}}$. Previous experiments show that $a$ and $b$
are typically at the level of 1\% and 0.01~MeV respectively. The
observed neutrino event number is proportional to the detector
volume, exposure time and reactor(s) power. A very powerful
reactor complex can consist of 8 reactor cores, each with
$\sim$3~GW thermal power. With a baseline of 58~km from such a
reactor complex and taking into account the oscillation
probability which is around $\sin^22\theta_{12}$, $5\times 10^5$
events, corresponding to a detector exposure of
$\sim$700~kt$\cdot$year, is taken as the default for an
experiment.

\begin{figure}[htbp]
\begin{center}
\subfigure[FCT spectrum.] 
{
    \includegraphics[width=0.4\textwidth]{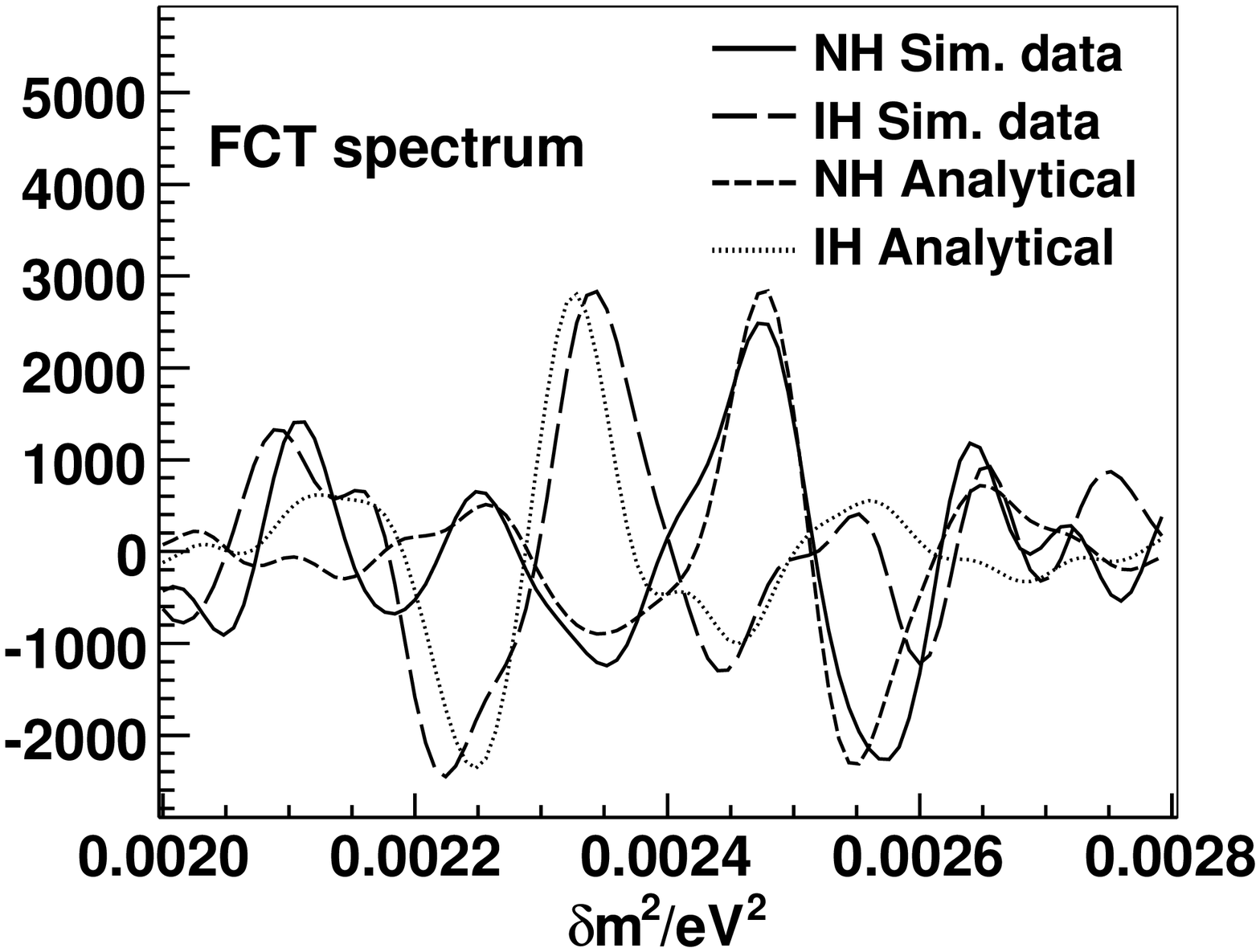}
}
\subfigure[FST spectrum.] 
{
    \includegraphics[width=0.4\textwidth]{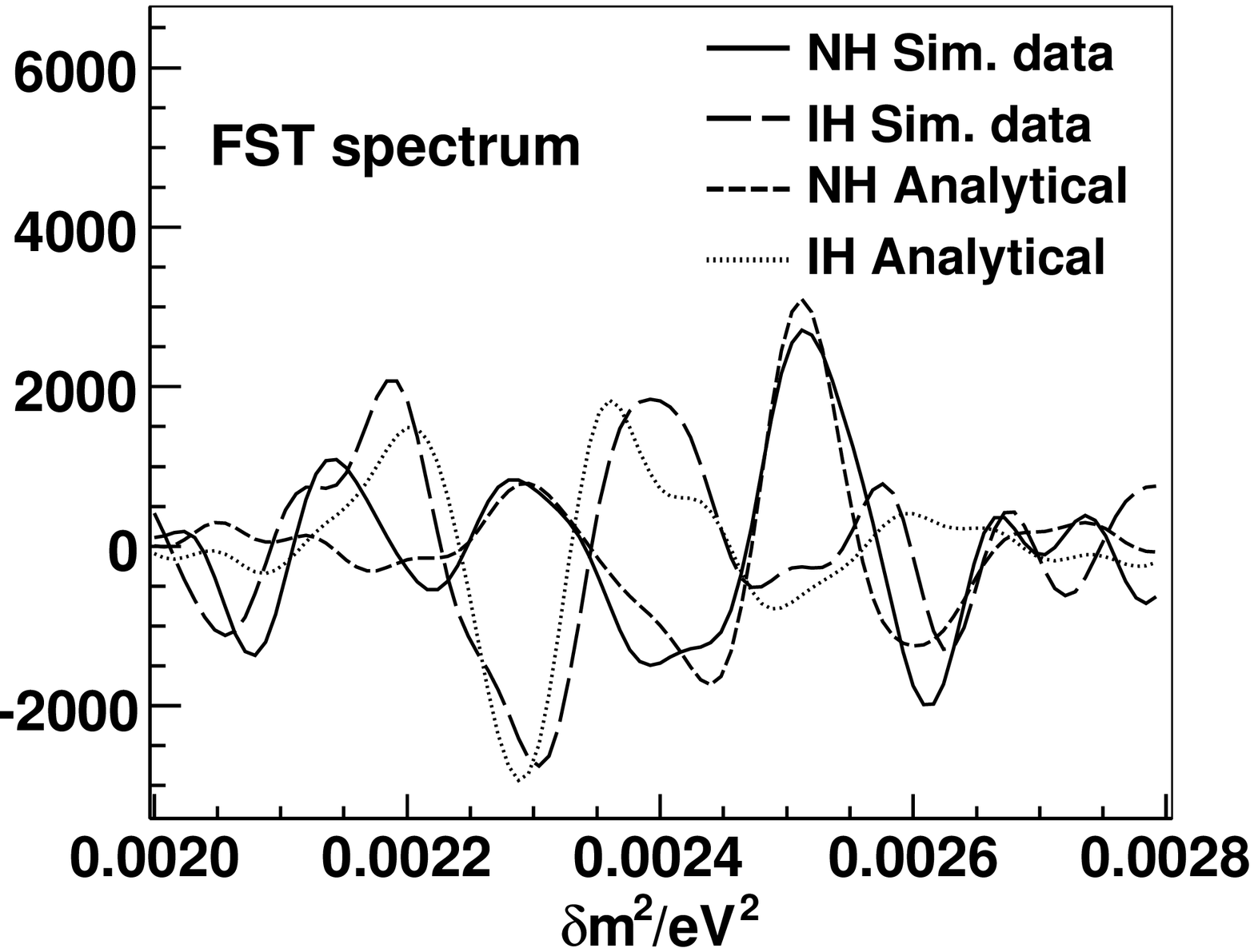}
}
\caption{
  FCT and FST spectra from simulation with parameters
  $(\sin^2(2\theta_{13}), \sigma_{E})$=$(0.02,3\%/\sqrt{E_{vis}})$,
  together with the analytical spectra for $\sin^2(2\theta_{13})=0.02$. Solid and
  long-dashed lines are spectra based on simulation for NH and IH
  cases, while dashed and dotted lines are analytical spectra.
   }
\label{fig:FTspec} 
\end{center}
\end{figure}

Fig.~\ref{fig:FTspec} shows FCT and FST spectra from a Monte Carlo
simulation using parameters $(\sin^2(2\theta_{13}),
\sigma_{E})=(0.02, 3\%/\sqrt{E_{vis}})$. For comparison, the
analytical spectra are also shown at $\sin^2(2\theta_{13}) =
0.02$. The impacts of the energy resolution and statistical errors
are obviously seen as that the amplitudes of noisy peaks and
valleys appear to be higher in the frequency range of $2.0\times
10^{-3} {\rm eV^2}<\Delta m^2<2.8\times 10^{-3} {\rm eV^2}$.
However, the main peak and valley are distinctive and can still be
used to determine the neutrino mass hierarchy.

We introduce parameters RL and PV \cite{Zhan:2008id} to quantify
the features of FCT and FST spectra.
\begin{eqnarray}
\label{eq:defRLPV}
 RL = \frac{RV-LV}{RV+LV},~ PV = \frac{P-V}{P+V}
\end{eqnarray}
where RV is the amplitude of the right valley and LV is that of
the left valley in the FCT spectrum; P is the amplitude of the
peak and V is that of the valley in the FST spectrum.

For each set of input parameters $\{\sin^2\theta_{13}, L,
\sigma_{E}, a, b, N\}$, we simulate 500 experiments and calculate
the probability to determine the mass hierarchy based on the
distributions of RL and PV values. The procedure is concluded as
the following:
\begin{enumerate}

\item Given $\sin^2\theta_{13}$ and $L$, we sample $N$ neutrino
events with energy ${E_{i}}(i=1,2,...,N)$ from energy spectrum
both for NH and IH cases.

\item $E_{i}$ is smeared and/or shifted to $E'_{i}$ based on the
given energy resolution ($\sigma_{E}$) and energy scale
uncertainty ($a$ and $b$) parameters.

\item FCT and FST spectra are calculated using Eq.~\ref{eq:FTexp}.

\item RL and PV values are calculated based on FCT and FST spectra
using Eq.~\ref{eq:defRLPV}.

\item Repeat the above steps 500 times and
obtain the distributions of RL and PV values.

\item Calculate the probability to determine the mass hierarchy
correctly based on the distributions of RL and PV values.

\end{enumerate}

\begin{figure}[htbp]
\begin{center}
\includegraphics[width=0.4\textwidth]{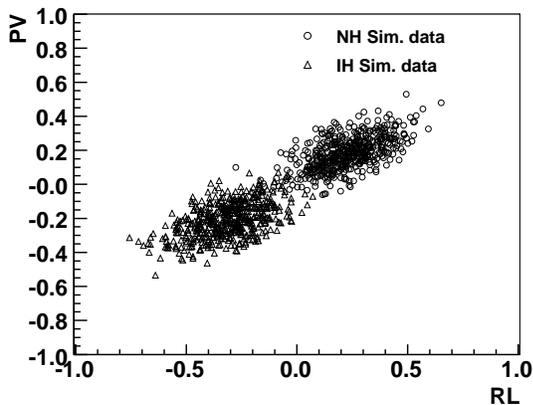}
\caption{Distribution of RL and PV values from 500 simulated
experiments with parameters
$(\sin^2(2\theta_{13}),\sigma_{E})=(0.02,2\%/\sqrt{E_{vis}})$. Two
clusters of points show the sensitivity to determine the mass
hierarchy.} \label{fig:RLPVexp}
\end{center}
\end{figure}

\begin{figure}[htbp]
\begin{center}
\includegraphics[width=0.48\textwidth]{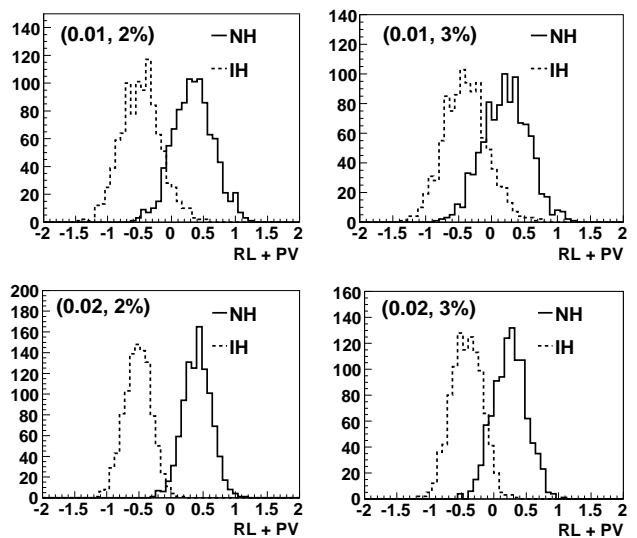}
\caption{Distribution of $RL+PV$ values from 1000 experiments for
parameters $(\sin^2(2\theta_{13}), \sigma_{E})$ being
$(0.01,2\%/\sqrt{E_{vis}})$, $(0.01,3\%/\sqrt{E_{vis}})$,
$(0.02,2\%/\sqrt{E_{vis}})$ and $(0.02,3\%/\sqrt{E_{vis}})$. Two
Gaussian distributions show the sensitivity to determine the mass
hierarchy.} \label{fig:RL+PVexp}
\end{center}
\end{figure}

Fig.~\ref{fig:RLPVexp} shows the distribution of RL and PV values
for 500 experiments with input parameters $(\sin^2(2\theta_{13}),
\sigma_{E})=(0.02,2\%/\sqrt{E_{vis}})$. Two clusters of points in
the (RL, PV) plane corresponding to NH and IH cases show the
probability  to determine the mass hierarchy. Various input
parameters have been tried and the distribution of $RL+PV$ is
shown in Fig.~\ref{fig:RL+PVexp}. Two clusters of points turn into
two Gaussian distributions and the probability to determine the
mass hierarchy can be correctly calculated.

\begin{figure}[htbp]
\begin{center}
\includegraphics[width=0.45\textwidth]{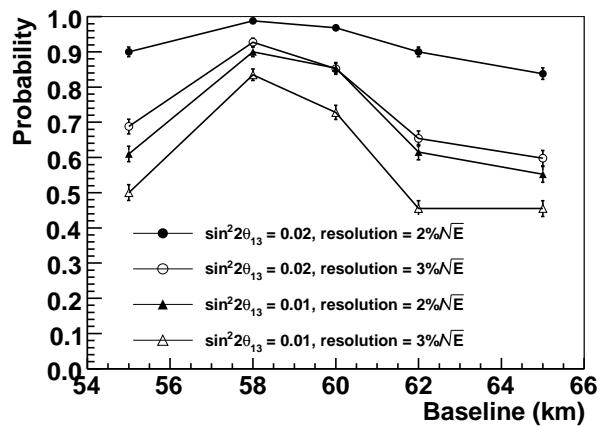}
\caption{Impact of the baseline to the determination probability
for 4 sets of parameters $(\sin^2(2\theta_{13}), \sigma_{E})$.}
\label{fig:prob_bl}
\end{center}
\end{figure}

To study the impact of the baseline, a total of 500 experiments
have been simulated for each set of input parameters
$(\sin^2(2\theta_{13}), \sigma_{E})$. Fig.~\ref{fig:prob_bl} shows
the results. The error bars are due to statistics since only a
limited number of experiments are simulated. The optimal baseline
is clearly 58~km, which is chosen as the default baseline.

\begin{figure}[htbp]
\begin{center}
\includegraphics[width=0.45\textwidth]{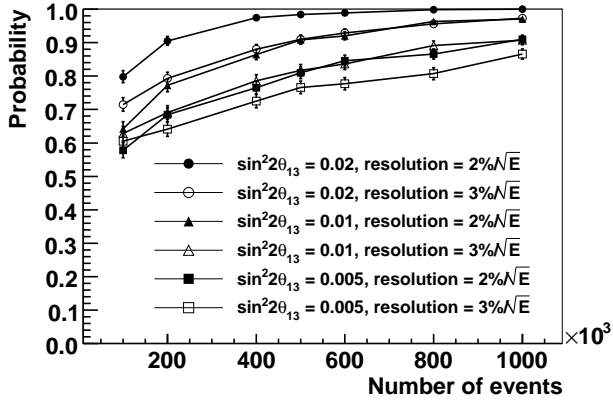}
\caption{Impact of the event number to the determination
probability in six sets of parameters $(\sin^2(2\theta_{13}),
\sigma_{E})$ } \label{fig:prob_N}
\end{center}
\end{figure}

Fig.~\ref{fig:prob_N} shows the impact of event number to the
determination probability. Obviously, fewer number of events will
induce larger statistical fluctuations, more noisy peaks and
valleys in the FCT and FST spectra and hence reduce determination
probability. As shown in Fig.~\ref{fig:prob_N}, a total of
$5\times 10^5$ events will reach $90\%$ determination probability
for $\sin^2(2\theta_{13})$=0.02 with an energy resolution of
$3\%/\sqrt{E_{vis}}$. This number of events, probably the largest
can be imagined nowadays, is chosen as the default in this paper.

\begin{figure}[htbp]
\begin{center}
\includegraphics[width=0.4\textwidth]{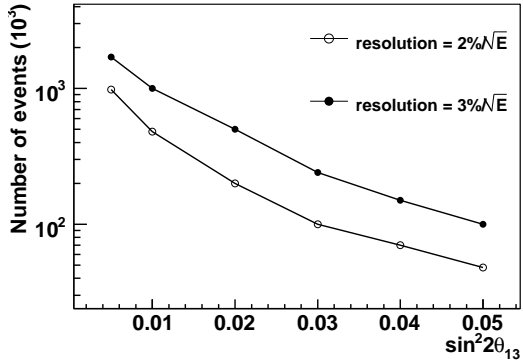}
\caption{Requirements to the number of events to determine the
mass hierarchy at $90\%$ probability as a function of
$\sin^2(2\theta_{13})$.} \label{fig:Nth}
\end{center}
\end{figure}

The requirement to the event statistics strongly depends on the
value of $\sin^2(2\theta_{13})$. Fig~\ref{fig:Nth} shows the
number of neutrino events needed to determine the mass hierarchy
at the $90\%$ confidence level as a function of
$\sin^2(2\theta_{13})$. Two cases of the energy resolution,
$2\%/\sqrt{E_{vis}}$ and $3\%/\sqrt{E_{vis}}$, are studied. If
$\sin^2(2\theta_{13})$ happens to be more than 0.05, as some of
the recent global fit indicated~\cite{Fogli:2008jx}, the number of
events can be a factor of 5 smaller than that in the case of
$\sin^2(2\theta_{13})$=0.02.

\begin{figure}[htbp]
\begin{center}
\includegraphics[width=0.4\textwidth]{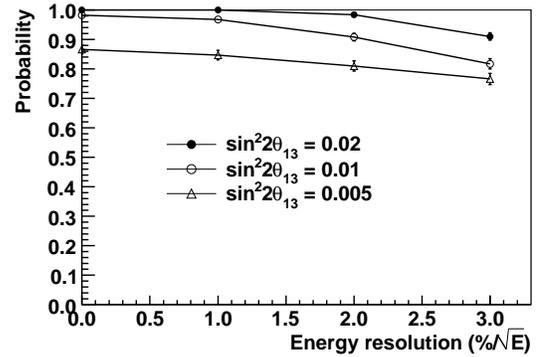}
\caption{Impact of the energy resolution to the determination
probability for $\sin^2(2\theta_{13})=0.02$, $0.01$ and $0.005$ }
\label{fig:prob_res}
\end{center}
\end{figure}

Impact of the energy resolution to the mass hierarchy
determination is studied for the cases of
$\sin^2(2\theta_{13})=0.02$, $0.01$ and $0.005$ as shown in
Fig.~\ref{fig:prob_res}. To achieve the mass hierarchy
determination probability better than $90\%$ at
$\sin^2(2\theta_{13})=0.02$, the energy resolution shall be better
than $3\%\sqrt{E_{vis}}$. This is actually a very stringent
requirement, at least a factor of two better than that of the
existing reactor neutrino experiments. For a typical liquid
scintillator experiment, substantial more light shall be collected
to reach such a level.

\begin{figure}[htbp]
\begin{center}
\includegraphics[width=0.4\textwidth]{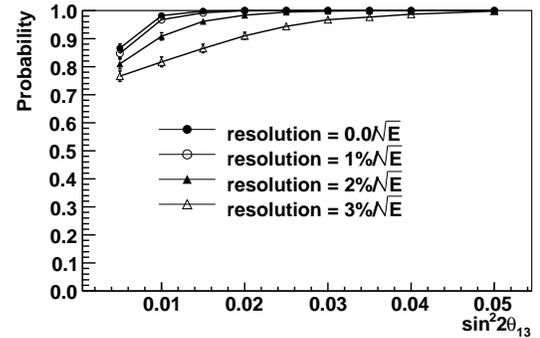}
\caption{Impact of $\sin^2(2\theta_{13})$ to the determination
probability in four cases of the energy resolution.}
\label{fig:prob_th}
\end{center}
\end{figure}

Fig.~\ref{fig:prob_th} shows the impact of $\sin^2(2\theta_{13})$
to the determination probability in four cases of the energy
resolution
\{$0/\sqrt{E_{vis}}$,~$1\%/\sqrt{E_{vis}}$,~$2\%/\sqrt{E_{vis}}$,~$3\%/\sqrt{E_{vis}}$\}.
It shows that the energy resolution is very important to determine
the mass hierarchy, while a larger $\sin^2(2\theta_{13})$ can
relax substantially such a requirement.

\begin{figure}[htbp]
\begin{center}
\includegraphics[width=0.4\textwidth]{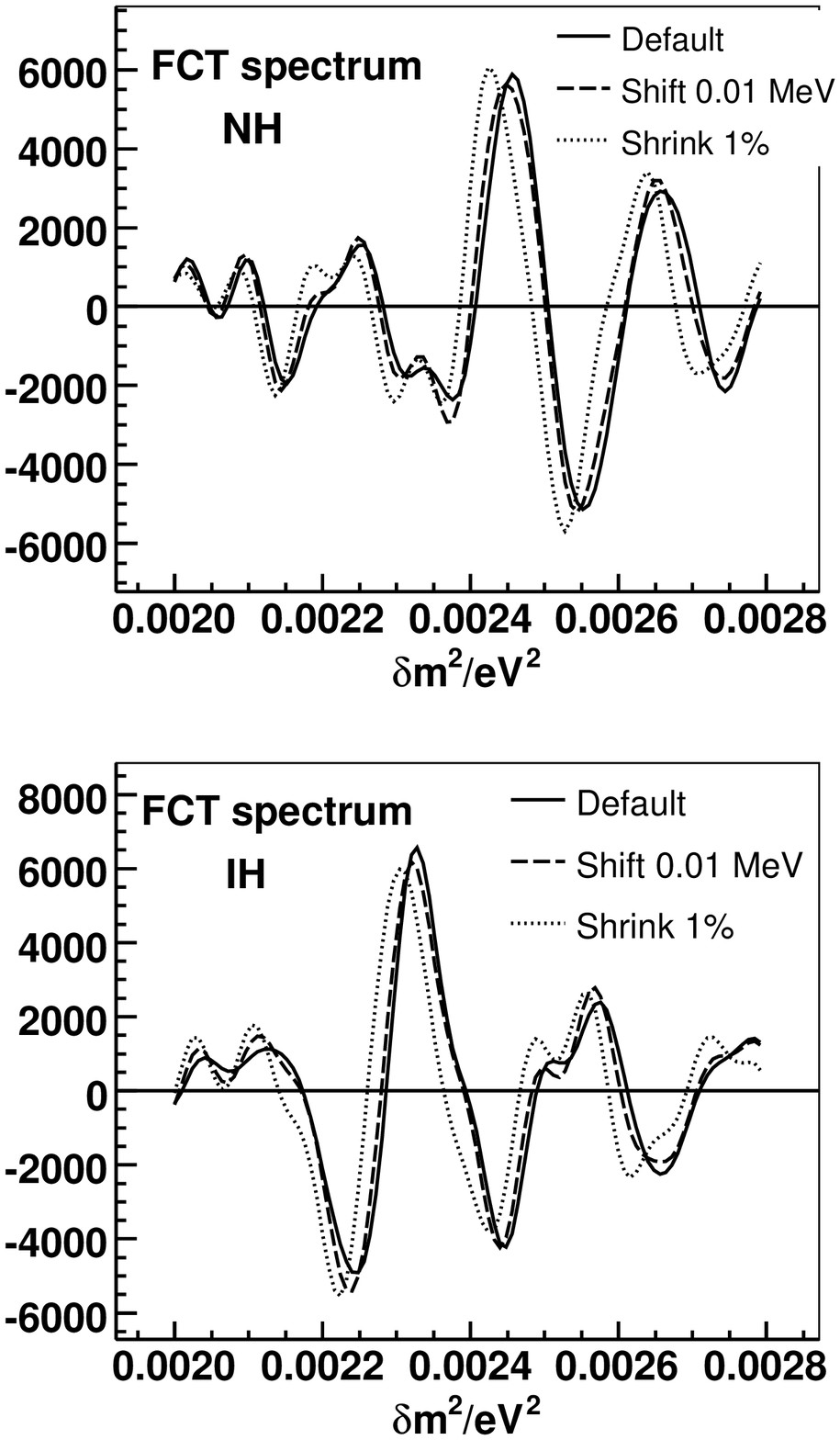}
\caption{FCT spectram for an energy shift of $0.01$ MeV and
shrinking of $1\%$.} \label{fig:Escale_minus_FT}
\end{center}
\end{figure}

The impact of the energy scale uncertainty is studied by
transforming the sampled neutrino energy $E$ to $E'=(1+a)E+b$. For
two cases of $a=-1\%$ or $b = -0.01$~MeV, which correspond to the
shrinking or left-shift of the neutrino energy spectrum, the FCT
spectra are calculated and shown in
Fig.~\ref{fig:Escale_minus_FT}. It shows that the FCT spectra,
both for NH and IH cases, are left shifted. After shrinking the
energy spectrum, the L/E spectrum expands and the oscillation
frequency becomes smaller, which results in the frequency spectra
(FCT spectra) left shifted. The energy scale uncertainty only
introduces a bias to the oscillation frequency and hence $\Delta
m^2_{31}$ (shown as the main peak in the FCT spectrum). Since our
method only depends on the relative position of peaks or valleys
in FCT and FST spectra, the mass hierarchy determination is not
affected by the energy scale uncertainty.

In summary, we have studied experimental requirements to determine
the mass hierarchy using Fourier cosine and sine transform to the
reactor neutrino L/E spectrum. The parameters RL and PV are
defined to extract features of the Fourier sine and cosine
spectra, and the mass hierarchy can be determined from events
collected in experiments similar to that in the analytical case.
The impacts of baseline, event statistics, energy resolution and
energy scale uncertainty to the mass hierarchy determination are
studied in detail. This paper provides a guidance to the design of
the experiment to determine the mass hierarchy using reactor
neutrinos.

\end{document}